# Study on Improving Microwave Heating Uniformity Based on Phase-Frequency Simultaneous Modulation Technique

Xu Zhu, Shaoyue Wang, Da He, Liping Yan, Jianan Hu, and Changjun Liu, *Senior Member*, *IEEE*

*Abstract*—Conventional microwave heating techniques are limited due to inherent thermal point residency effects and inadequate control over the heating process. A novel method is proposed to enhance microwave heating uniformity using the injection-pulling technique. In this method, the injection-pulling technique is employed to achieve simultaneous modulation of both the output phase and frequency of the magnetron, thereby extending the locking bandwidth of the injection-locking technique. The output characteristics of the injection-pulled magnetron were validated through numerical calculations and experiments. Microwave heating experiments were conducted under both a five-cup water load and an absorbent paper load. Compared to conventional injection-locking frequency sweeping, the proposed method not only expands the sweeping bandwidth from 8 MHz to 18 MHz but also further improves heating uniformity, offering more options for magnetron applications in microwave heating.

## I. INTRODUCTION

WITH the advancement of electronic technology, microwave heating has gained increasing attention in industrial applications, such as food processing and plastics manufacturing [1][2][3]. Magnetrons, which offer the advantages of low cost, compact size, high output power, and high efficiency, have become the predominant choice for microwave heating in industrial applications [4][5]. However, a magnetron demonstrates inherent limitations including power output instability and poor phase noise characteristics [6][7]. Consequently, microwave heating systems are frequently prone to issues such as non-uniform heating, load mismatch, and inadequate process controllability during operational heating procedures [8][9].

Extensive research has been conducted to solve the problem of poor microwave heating uniformity [10][11]. The most common method for improving the uniformity of microwave heating is to use mode stirrers, which enhance uniformity by altering the electromagnetic field modes in the cavity [12][13][14][15]. Additionally, multi-frequency heating is proposed to address the issues associated with single-frequency operation. Yang et al. proposed a frequency-complementary technique to enhance the uniformity of microwave heating [16]. Du et al. validated that multi-frequency excitation within 3.6-10 MHz frequency intervals could not only improve heating uniformity but also, to some extent, enhance heating efficiency [17].

The injection-locking technique is also a valid method for achieving variable-frequency heating, which provides precise control of a magnetron operational frequency and phase through the injection of an external signal with proximity to the magnetron free-running frequency [18][19][20]. Yang et al. implemented dynamic frequency sweeping with a 20-MHz bandwidth during heating processing, which significantly improved heating uniformity [21]. However, extant research primarily focuses on the injection-locking state of a magnetron. The spectrum outside the locking bandwidth is not effectively utilized, which limits the selectable frequency range of the sweep.

This paper proposes a novel method to enhance heating uniformity by leveraging the injection-pulling state of the magnetron. This methodology utilizes a single-sweep frequency signal to achieve simultaneous modulation of both the output phase and frequency of the magnetron. Microwave heating experiments were conducted using various modulation strategies, including sweep frequency within the injection-pulling range. The results demonstrate an improvement in heating uniformity for the five-cup water load and absorbent paper load.

## II. INJECTION-PULLING MAGNETRONS

Define the angular frequency of the resonant cavity as $\omega_0$, when a low-power external signal at a frequency around $\omega_0$ is injected, it can be equivalent to a parallel load of the magnetron. Upon considering the temporal evolution, the circuit equation can be decomposed into a pair of normalized, slowly time-varying differential equations [22]:

$$\frac{d\theta}{dt} + 1 - \omega_1 = \frac{\rho}{2Q_{ext}}\sin\theta \tag{1}$$

$$\frac{dV_{RF}}{dt} + \frac{V_{RF}}{Q_0}\left(\frac{1}{1} - \frac{1}{V_{RF}}\right) = -V_{RF}\frac{\rho}{2Q_{ext}}\cos\theta \tag{2}$$

This work was supported in part by the National Natural Science Foundation of China under Grants U2015A22 and 60301004.

X. Zhu, S. Wang, D. He, and C. Liu are with the School of Electronics and Information Engineering, Sichuan University, Chengdu 610064, China (*Corresponding author: Changjun Liu.* e-mail: cjliu@ieee.org).

J. Hu is with the Shenzhen Technology Center of Guangdong Whirlpool Electrical Appliances Co. Ltd., Shenzhen 518000, China

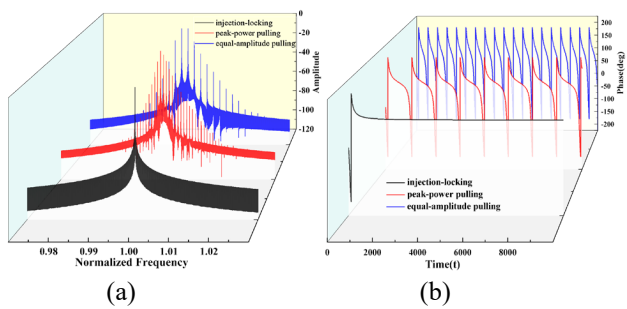

**Fig. 1.** The (a) spectrum and (b) time-varying phase of magnetron under various states.

where $Q_0$ and $Q_{ext}$ are the unloaded quality factor and external quality factor of the resonant circuit, $\theta$ is the phase difference between the external signal and magnetron output, and $V_{RF}$ is the instantaneous microwave output amplitude. All frequencies are normalized by $\omega_0$, where $\omega_I$ is the normalized angular frequency of the injected signal, and $\rho = \sqrt{P_1/P_0}$ is the injection ratio. (1) describes the phase change rate of the magnetron when an external signal is injected. For the magnetron to maintain the steady state, $d\theta/dt$ must be zero. Then (2) is described by the well-known Adler locking equation [23]:

$$|1 - \omega_I| \leq \rho/2Q_{ext} \qquad (3)$$

Here, $\rho/2Q_{ext}$ is defined as the injection intensity, denoted as $k$. When the injected signal frequency exceeds the range defined by (3), the magnetron will exhibit a multi-frequency output state. A magnetron in this state is often considered to be unlocked. Scholars refer to it as the injection-pulling state [24].

For a more direct understanding of this state, the output power spectrum and time-varying phase of the magnetron were calculated, (1) and (2) were solved using the fourth-order Runge-Kutta method, and the output spectrum was calculated via the Fast Fourier Transformation (FFT). In the calculation, $Q_0$ is set to 1200, and k is set to 0.002.

Fig. 1(a) illustrates the calculated output spectrum of the magnetron under injection-locking and injection-pulling states. The injection-pulled magnetron's output spectrum introduces additional frequency components compared to the injection-locking state. When the injected signal's frequency just exceeds the locking range, the frequency of the magnetron's peak power is still controlled by the injected signal, and this state is referred to as peak-power pulling. As the frequency of the injected signal continues to deviate, the amplitude of the magnetron's primary peak gradually diminishes. In contrast, the secondary peak's amplitude progressively grows, eventually evolving into a state of equal-amplitude output. Therefore, based on the characteristics of the injection-pulling state, if a sweep frequency signal is introduced, the controllable range of its peak power has been enhanced compared to conventional injection locking.

Furthermore, as shown in Fig. 1(b), the injection-pulled magnetron's output phase contains a modulation component, and it undergoes periodic temporal variations. This phenomenon indicates that complex modulation of the magnetron's frequency and phase is incorporated into the output. This time-varying frequency and phase result in rapid fluctuations of the microwave modes within the heating cavity, thereby enhancing the uniformity of heating. Microwave heating experiments were conducted using this method to verify the improvement in heating uniformity.

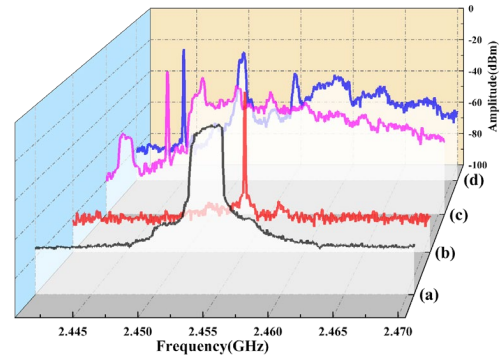

**Fig. 2.** Magnetron output spectra with (a) free-running, (b) injection-locking, (c) peak-power pulling, and (d) equal-amplitude pulling states.

## III. Microwave Heating Experiments

### A. Experimental Preparation

Under experimental conditions, the magnetron's output exhibits four distinct states as illustrated in Fig. 2: free-running, injection-locking, peak-power pulling, and equal-amplitude pulling states. At the peak-power pulling and equal-amplitude pulling states, their spectra are shown in Fig. 2(c) and (d) with the injection locking frequency lower than the locking band. Similar spectra will appear when the injection locking frequency is higher than the locking band. The following modulation strategies have been formulated: I. Free-running state; II. Sweeping frequency within the injection-locking range (2.450 – 2.458 GHz); and III. Sweeping frequency within the injection-pulling and injection-locking range (2.443 – 2.461 GHz).

An experimental system is established to verify the heating performance of the injection-modulated magnetron. The diagram of the magnetron experimental system used in this paper is shown in Fig. 3(a). Fig. 3(b) shows a photograph of the system. The magnetron (LG 2M22603GWH) in the experiment is powered by a switch-mode power supply (WELAMP 2000F, Magmeet). The output spectrum of the magnetron is measured by a spectrum analyzer (FSV, R&S), and the output power is measured by a power meter (AV2433, Ceyear). A signal generator (HMC-T2220, Hittite) and a power amplifier (ZHL-30W-262, Mini-Circuits) generate the injection signal. We utilize a host computer to control the sweeping frequency of the signal generator. We use two circulators to inject an external signal, thereby protecting the power amplifier and signal generator. The experimental procedure is as follows:

Five cups, each containing 100g ± 1g of pure water, are placed on a thermally insulating pad, and the initial temperature is measured using an UNI-T UT325F thermometer. The cups are positioned in the cavity according to the configuration in Fig. 3(b), and heating is maintained for 95 seconds. After the heating process, remove the cups and return them to the insulating pad.

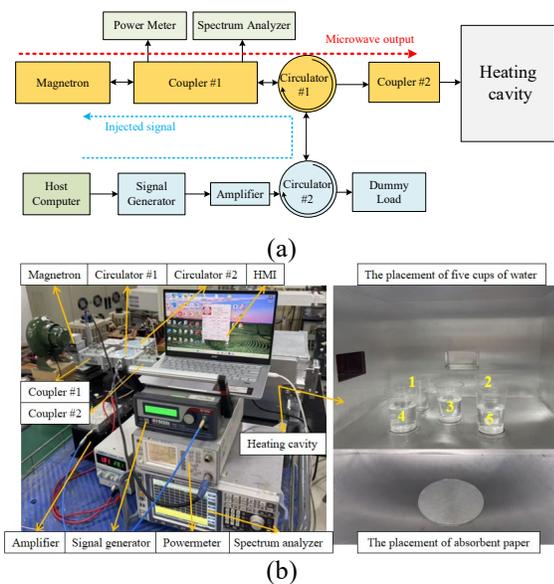

(a)

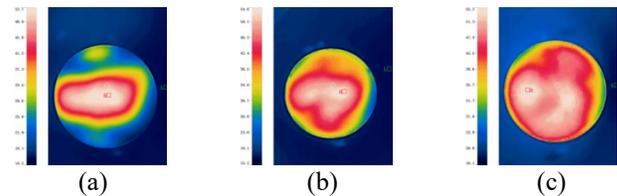

(b)

**Fig. 3.** Setup of the experiment system. (a) Diagram of the experiment system. (b) Photograph of the experiment system

We sequentially stir the water in each cup for three seconds to ensure uniform temperature, and record the final temperature of each sample. To minimize the impact of heat dissipation during stirring, the heating process is repeated, with water in each cup stirred in reverse order and the final temperature recorded.

The uniformity of the results is reflected using the coefficient of variation (*COV*) of the temperature:

$$COV = \sigma / (T_a - T_0) \tag{4}$$

Here, $\sigma$ is the standard deviation of the overall temperature, while $T_a$ and $T_0$ are the average temperatures after and before heating. We calculate the *COV* using the average temperature in forward and reverse heating experiments. A lower *COV* indicates an improved heating uniformity.

In addition to the five-cup load, using BJM 811 qualitative filter paper with a diameter of 12 cm saturated with pure water as another load, to investigate the impact of heating strategies on various loads. The absorbent paper is positioned at the center of the cavity, as shown in Fig. 3(b), and subsequently heated for 25 seconds.

### B. Results and Analysis

The heating results of the absorbent paper are pictured by the UNI-T UTi260B professional thermal imager, as shown in Fig. 4. For each group, 160 hot spots are selected, and their *COVs* are computed. The heating results under different loads are shown in TABLE I.

For both the five-cup water load and the absorbent paper load, the heating uniformity results consistently show strategy III > II > I. Compared to the free-running condition, the heating uniformity of a five-cup water load improved by 9% under strategy III. In contrast, the uniformity improvement for the absorbent paper load was even more pronounced. Furthermore, the improvement in hotspot distribution across different strategy is more visually demonstrated in Fig. 4.

In the free-running state, the magnetron operates at a single fixed frequency, resulting in the poorest heating uniformity.

(a)        (b)        (c)

**Fig. 4.** Results of absorbent paper under different heating strategies: (a) strategy I, (b) strategy II, and (c) strategy III.

Under conventional injection-locking frequency sweeping, the magnetron output is precisely locked by an external signal with phase coherence, featuring a narrow frequency sweep range and relatively slow, orderly variations. After introducing the injection-pulling state, the magnetron output exhibits multiple-frequency components, intense phase jumps, strong randomness, and a broader frequency sweep range, ultimately resulting in improved heating uniformity.

TABLE II compares the proposed method with the conventional injection-locking frequency sweeping method. It is observed that under identical injection intensity, the proposed method extends the frequency sweep bandwidth from 8 MHz to 18 MHz, providing additional spectral range for sweep strategies. Moreover, the experiments achieved multi-frequency scanning output with phase randomness using only a single external sweep frequency signal.

### TABLE I
### HEATING RESULTS OF DIFFERENT LOADS

| Load | Strategy I | Strategy II | Strategy III |
|---|---|---|---|
| FIVE-CUP WATER | 0.365 | 0.348 | 0.332 |
| ABSORBENT PAPER | 0.437 | 0.218 | 0.135 |

### TABLE II
### COMPARISON OF DIFFERENT METHODS

| Strategy | Bandwidth | Output Phase | Output Frequency |
|---|---|---|---|
| Conventional injection-locking | 8MHz | Stable | Single-frequency |
| Proposed method | 18MHz | Random | Multi-frequency |

### IV. CONCLUSION

This paper elucidates the output characteristics of the magnetron in the injection-pulling state through numerical simulations. It proposes a novel method to achieve simultaneous phase-frequency modulation of the magnetron. Subsequently, microwave heating experiments were conducted. Experimental results demonstrate that, under different loads, the sweep-frequency scheme incorporating the injection-pulling state not only extends the bandwidth of conventional injection-locking frequency sweeping schemes but also delivers substantial uniformity enhancement, offering a novel perspective on applying injection modulation magnetrons in the microwave heating field.